\documentclass[]{article}
\usepackage{lmodern}
\usepackage{amssymb,amsmath}
\usepackage{ifxetex,ifluatex}
\usepackage{fixltx2e} 
\ifnum 0\ifxetex 1\fi\ifluatex 1\fi=0 
  \usepackage[T1]{fontenc}
  \usepackage[utf8]{inputenc}
\else 
  \ifxetex
    \usepackage{mathspec}
  \else
    \usepackage{fontspec}
  \fi
  \defaultfontfeatures{Ligatures=TeX,Scale=MatchLowercase}
\fi
\IfFileExists{upquote.sty}{\usepackage{upquote}}{}
\IfFileExists{microtype.sty}{%
\usepackage{microtype}
\UseMicrotypeSet[protrusion]{basicmath} 
}{}
\usepackage[margin=1in]{geometry}
\usepackage{hyperref}
\hypersetup{unicode=true,
            pdftitle={regsem: Regularized Structural Equation Modeling},
            pdfauthor={Ross Jacobucci; University of Notre Dame},
            pdfborder={0 0 0},
            breaklinks=true}
\usepackage{natbib}
\usepackage{color}
\usepackage{fancyvrb}

\DefineVerbatimEnvironment{Highlighting}{Verbatim}{commandchars=\\\{\}}
\usepackage{framed}
\definecolor{shadecolor}{RGB}{248,248,248}
\newenvironment{Shaded}{\begin{snugshade}}{\end{snugshade}}
\newcommand{\KeywordTok}[1]{\textcolor[rgb]{0.13,0.29,0.53}{\textbf{{#1}}}}
\newcommand{\DataTypeTok}[1]{\textcolor[rgb]{0.13,0.29,0.53}{{#1}}}
\newcommand{\DecValTok}[1]{\textcolor[rgb]{0.00,0.00,0.81}{{#1}}}

\newcommand{\StringTok}[1]{\textcolor[rgb]{0.31,0.60,0.02}{{#1}}}

\newcommand{\CommentTok}[1]{\textcolor[rgb]{0.56,0.35,0.01}{\textit{{#1}}}}

\newcommand{\NormalTok}[1]{{#1}}
\usepackage{graphicx,grffile}
\makeatletter
\def\maxwidth{\ifdim\Gin@nat@width>\linewidth\linewidth\else\Gin@nat@width\fi}
\def\maxheight{\ifdim\Gin@nat@height>\textheight\textheight\else\Gin@nat@height\fi}
\makeatother
\setkeys{Gin}{width=\maxwidth,height=\maxheight,keepaspectratio}
\IfFileExists{parskip.sty}{%
\usepackage{parskip}
}{
\setlength{\parindent}{0pt}
\setlength{\parskip}{6pt plus 2pt minus 1pt}
}
\ifx\paragraph\undefined\else
\let\oldparagraph\paragraph
\renewcommand{\paragraph}[1]{\oldparagraph{#1}\mbox{}}
\fi
\ifx\subparagraph\undefined\else
\let\oldsubparagraph\subparagraph
\renewcommand{\subparagraph}[1]{\oldsubparagraph{#1}\mbox{}}
\fi

\let\rmarkdownfootnote\footnote%
\def\footnote{\protect\rmarkdownfootnote}

\usepackage{titling}

  \title{regsem: Regularized Structural Equation Modeling}
  \pretitle{\vspace{\droptitle}\centering\huge}
  \posttitle{\par}
  \author{Ross Jacobucci \\ University of Notre Dame}
  \preauthor{\centering\large}
  \postauthor{\par}
  \date{}
  \predate{}\postdate{}

\begin{document}
\maketitle
\begin{abstract}
The \textit{regsem} package in \textit{R}, an implementation of
regularized structural equation modeling
\citep[RegSEM;][]{jacobucci2016regularized}, was recently developed with
the goal of incorporating various forms of penalized likelihood
estimation with a broad array of structural equations models. The forms
of regularization include both the \textit{ridge} \citep{hoerl1970} and
the least absolute shrinkage and selection operator
\citep[\textit{lasso};][]{Tibshirani1996}, along with extensions that
lead to sparser solutions. RegSEM is particularly useful for structural
equation models that have a small parameter to sample size ratio, as the
addition of penalties can reduce the complexity, thus reducing the bias
of the parameter estimates. The paper covers the algorithmic details and
an overview of the use of \textit{regsem} with the application of both
factor analysis and latent growth curve models.

\end{abstract}

\section{Introduction}\label{introduction}

In the context of latent variable, reducing the complexity of models can
come in many forms: selecting among multiple predictors of a latent
variable, simplifying factor structure by removing cross-loadings,
determining whether the addition of nonlinear terms are necessary in
longitudinal models, and many others. The aim of performing variable
selection in structural equation models could be motivated by either an
inadequate sample size (in relation to the number of parameters), or
simply to present a more parsimonious relationship between
variables.Particularly when the number of variables are high, reducing
the model complexity in a globally optimal way can be challenging.

As a simple running example, Figure 1 depicts a linear latent growth
curve model \citep[e.g.][]{meredith1990latent} with four time points and
ten predictors for a simulated dataset. In this, a researcher may want
to test this model, but may only have a relatively small sample size
(e.g.~80). There are 29 estimated parameters in this model, resulting in
a estimated parameter to sample size ratio far below even the most
liberal recommendations \citep[e.g.~10:1 parameters to sample
size;][]{kline2015principles}. In lieu of finding additional
respondents, reducing the number of parameters estimated is one
effective strategy for reducing bias. Specifically, the 20 estimated
regressions from \textit{c1-c10} could be reduced to a number that makes
the ratio of the parameters estimated to sample size more reasonable. To
explore this further, the next section provides an overview of
regularization, and how different forms can be used to perform variable
selection across a broad range of models.

\begin{figure}
    \centering
    \includegraphics[width=.5\linewidth]{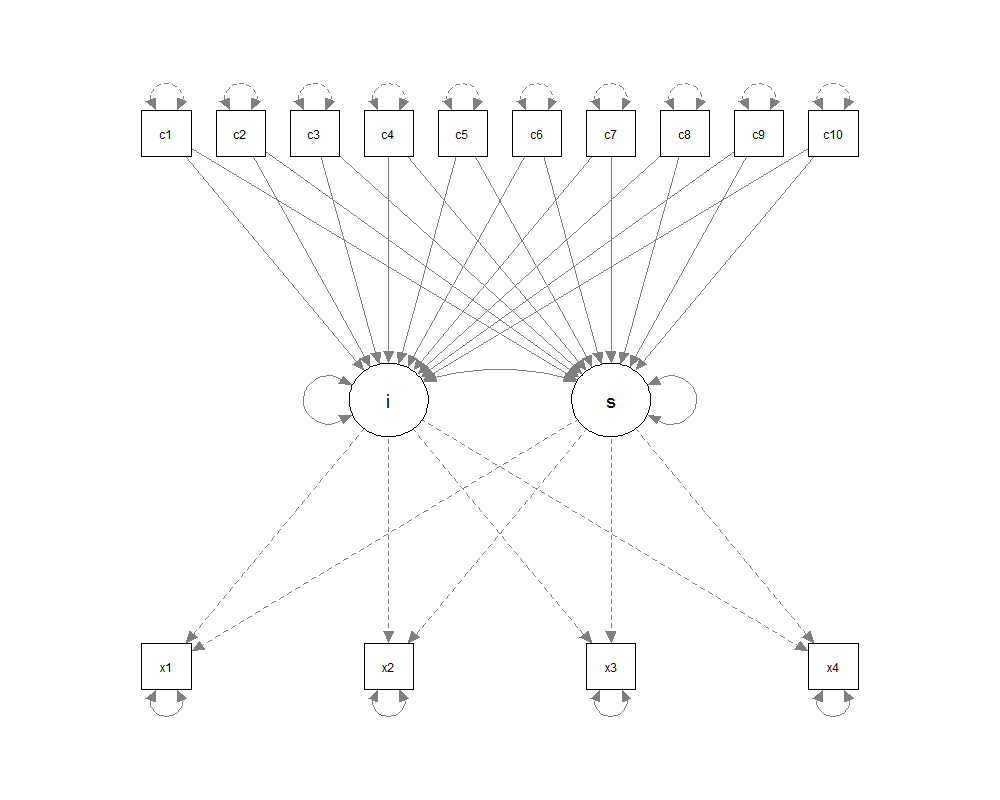}
    \caption{Growth curve model with 10 predictors of both the intercept and slope}
\end{figure}

\section{Regularization}\label{regularization}

Although a host of methods exist to perform variable selection, the use
of regularization has seen a wide array of application in the context of
regression, and more recently, in areas such as graphical modeling, as
well as a host of others. The two most common procedures for
regularization in regression are the \textit{ridge} \citep{hoerl1970}
and the least absolute shrinkage and selection operator
\citep[\textit{lasso};][]{Tibshirani1996}; however, there are various
alternative forms that can be seen as subsets or generalizations of
these two procedures. Given an outcome vector \textit{y} and predictor
matrix \(X \in {R}^{n \times p}\) , ridge estimates are defined as

\[\tag{1}
\hat{\beta}^{ridge}= argmin \Big\{ \sum_{i=1}^{N} (y_{i} = \beta_{0} - \sum_{j=1}^{p}x_{ij} \beta_{j})^{2}  + \lambda \sum_{j=1}^{p} \beta_{j}^{2}\Big\},
\]

where \(\beta_{0}\) is the intercept, \(\beta_{j}\) is the coefficient
for \(x_{j}\), and \(\lambda\) is the penalty that controls the amount
of shrinkage. Note that when \(\lambda = 0\), Equation 3 reduces to
ordinary least squares regression. As \(\lambda\) is increased, the
\(\beta\) parameters are shrunken towards zero. The lasso estimates are
defined as

\[\tag{2}
\hat{\beta}^{lasso}= argmin \Big\{ \sum_{i=1}^{N} (y_{i} = \beta_{0} - \sum_{j=1}^{p}x_{ij} \beta_{j})^{2}  + \lambda \sum_{j=1}^{p}|\beta_{j}|\Big\}.
\]

In lasso regression, the \(l_{1}\)-norm is used, instead of
\(l_{2}\)-norm as in ridge, which also shrinks the \(\beta\) parameters,
but additionally drives the parameters all the way to zero, thus
performing a form of subset selection.

In the context of our example depicted in Figure 1, to use lasso
regression to select among the covariates, the growth model would need
to be reduced to two factor scores, which neglects both the relationship
between both the slope and intercept, reducing both to independent
variables. Particularly in models with a greater number of latent
variables, this becomes increasingly problematic. A method that keeps
the model structure, while allowing for penalized estimation of specific
parameters is regularized structural equation modeling
\citep[RegSEM;][]{jacobucci2016regularized}. RegSEM adds a penalty
function to the traditional maximum likelihood estimation (MLE) for
structural equation models (SEMs). The maximum likelihood cost function
for SEMs can be written as

\[\tag{3}
F_{ML}=log(\left|\Sigma\right|)+tr(C*\Sigma^{-1})-log(\left|C\right|)- p.
\]

where \(\Sigma\) is the model implied covariance matrix, \(C\) is the
observed covariance matrix, and \(p\) is the number of estimated
parameters. RegSEM builds in an additional element to penalize certain
model parameters yielding

\[\tag{4}
F_{regsem} = F_{ML} + \lambda P(\cdot)
\]

where \(\lambda\) is the regularization parameter and takes on a value
between zero and infinity. When \(\lambda\) is zero, MLE is performed,
and when \(\lambda\) is infinity, all penalized parameters are shrunk to
zero. \(P(\cdot)\) is a general function for summing the values of one
or more of the model's parameter matrices. Two common forms of
\(P(\cdot)\) include both the lasso (\(\| \cdot \|_{1}\)), which
penalizes the sum of the absolute values of the parameters, and ridge
(\(\| \cdot \|_{2}\)), which penalizes the sum of the squared values of
the parameters.

In our example, the twenty regression parameters from the covariates to
both the intercept and slope would be penalized. Using lasso penalties,
the absolute value of these twenty parameters would be summed and after
being multiplied by the penalty \(\lambda\), added to equation 4,
resulting in:

\[\tag{5}
F_{lasso} = F_{ML} + \lambda * \left\|  \begin{matrix}  
c1_{i} \\
c2_{i}\\
\vdots \\
c10_{i}\\
c1_{s}\\
\vdots \\
c10_{s}\\
\end{matrix}  \right\|_{1}
\]

Although the fit of the model is easily calculated given a set of
parameter estimates, traditional optimization procedures for SEM cannot
be used given the non-differentiable nature of lasso penalties, and as
detailed later, sparse extensions.

\subsection{Optimization}\label{optimization}

One method that has become popular for optimizing penalized likelihood
method is that of proximal gradient descent \citep[e.g.~p.~104
in][]{hastie2015statistical}. In comparison to one-step procedures
common in SEM optimization, that only involve a method for calculating
the step size and the direction (typically using the gradient and an
approximation of the Hessian), proximal gradient descent can be
formulated as a two-step procedure. With a stepsize of \(s^{t}\) and
parameters \(\theta^{t}\) at iteration \textit{t}:

\begin{enumerate}
    \item First, take a gradient step size $z = \theta^{t} - s^{t} \nabla g(\theta^{t})$.
    \item Second, perform elementwise soft-thresholding $\theta^{t+1} = S_{s^{t} \lambda}(z)$.
\end{enumerate}

where \(S_{s^{t} \lambda}(z)\) is the soft-thresholding operator
\citep{donoho1995noising} used to overcome non-differentiability of the
lasso penalty at the origin:

\begin{equation}
S_{s^{t} \lambda}(z_{j}) = sign(\theta_{j})(|\theta_{j}|-s^{t} \lambda)_{+}.
\end{equation}

In this, \((x)_{+}\) is shorthand for max(x,0) and \(s^{t}\) is the step
size. Henceforth, \(\lambda\) is assumed to encompass both the penalty
and the step size \(s^{t}\). This procedure is only used to update
parameters that are subject to penalty. Non-penalized parameters are
updated only using step 1 from above.

However, in testing with larger SEMs, the use of only the gradient for
minimization has been found to cause problems. Particularly at higher
penalties, estimation of both observed and latent variances can become
difficult, as these parameter estimates can become inflated if the
optimization routine has a hard time finding an optima. Due to this, for
larger models it is recommended to use a quasi-Newton method,
specifically the BFGS (Broyden-Fletcher-Goldfarb-Shanno) algorithm. This
method involves computing approximations to the Hessian matrix of the
objective function, in which step 1 above is replaced with:

\begin{equation}
z = \theta^{t} - s^{t} H^{-1} \nabla g(\theta^{t}).
\end{equation}

Although calculating the approximation to the Hessian is computationally
intensive, this minimization method paired with a backtracking rule for
finding the step size (\(s^{t}\)) has been found to be only slightly
slower than gradient descent. However, more testing is needed to
determine more specifically which settings each of the optimization
methods may be preferential.

\section{Types of Penalties}\label{types-of-penalties}

Outside of both ridge and lasso penalties, a host of additional forms of
regularization exist.

\subsection{Elastic Net}\label{elastic-net}

Most notably, the elastic net \citep{zou2005regularization} encompasses
both the ridge and lasso, reaching a compromise between both through the
addition of an additional parameter \(\alpha\), manifesting itself as

\[
P_{enet}(\theta_{j}) = (1-\alpha)\| \theta_{j} \|_{2} + \alpha\| \theta_{j} \|_{1}
\]

with a soft-thresholding update of \[
S(\theta_{j})= 
\begin{cases}
0,&  |\theta_{j}| < \alpha\lambda\\
\frac{sgn(\theta_{j})(|\theta_{j}|-\alpha\lambda)}{1+(1-\alpha)\lambda},              & |\theta_{j}|\geq\alpha\lambda
\end{cases}
\]

When \(\alpha\) is zero, ridge is performed, and conversely when
\(\alpha\) is 1, lasso regularization is performed. This method
harnesses the benefits of both methods, particularly when variable
selection is warranted (lasso), but there may be collinearity between
the variables (ridge).

\subsection{Adaptive Lasso}\label{adaptive-lasso}

In using lasso penalties, difficulties emerge when the scale of
variables differ dramatically. By only using one value of \(\lambda\),
this can add appreciable bias to the resulting estimates
\citep[e.g.][]{fan2001variable}. One method proposed for overcoming this
limitation is the adaptive lasso \citep{zou2006adaptive}. Instead of
penalizing parameters directly, each parameter is scaled by the
un-penalized estimated (MLE parameter estimates in SEM). The adaptive
lasso results in: \[
F_{alasso} = F_{ML} + \lambda \| \theta_{ML}^{-1} * \theta_{pen} \|_{1}
\]

with, following the same form for the lasso, the soft-thresholding
update is:

\[
S(\theta_{j})= sign(\theta_{j})(|\theta_{j}|-\frac{\lambda}{2|\theta_{j}|})_{+}
\]

In this, larger penalties are given for non-significant (smaller)
parameters, limiting the bias in estimating larger, significant
parameters. Note that one limitation of this approach for SEM models is
that the model needs to be estimable with MLE. Particularly for models
with large numbers of variables, in relation to sample size, this may
not be possible.

\subsection{Sparse Extensions}\label{sparse-extensions}

\begin{figure}
    \centering
    \includegraphics[width=.5\linewidth]{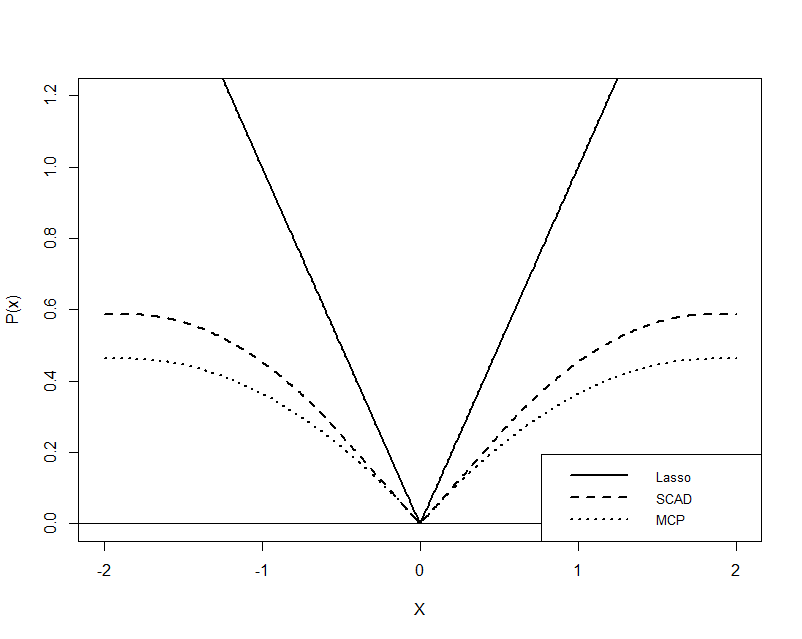}
    \caption{Comparison of types of penalties with $\lambda=0.5$}
\end{figure}

Two additional penalties that overcome some of the deficiencies of the
lasso, producing sparser solutions, include the smoothly clipped
absolute deviation penalty \citep[SCAD;][]{fan2001variable} and the
minimax concave penalty \citep[MCP;][]{zhang2010nearly}. In comparison
to the lasso, both the SCAD and MCP have much smaller penalties for
large parameters, where the amount of penalty for small penalties is
similar to the lasso, as is evident in Figure 2.

The SCAD takes the form of:

\[
pen_{\lambda,\gamma}(\theta_{j}) = \lambda \big\{I(\theta_{j}\leq\lambda) + \frac{(\gamma \lambda-0)_{+}}{(\gamma-1)\lambda}I(\theta_{j}>\lambda)\big\}
\]

with a soft-thresholding update of \[
S(\theta_{j})= 
\begin{cases}
S(\theta_{j},\lambda),&  |\theta_{j}| \geq 2\lambda\\
\frac{\gamma-1}{\gamma-2}S(\theta_{j},\frac{\lambda\gamma}{\gamma-1}),              & 2\lambda < |\theta_{j}|\leq\alpha\lambda\\
\theta_{j} & |\theta_{j}| > \lambda \gamma
\end{cases}
\]

for \(\gamma > 2\). As the the penalty in equation 11 is non-convex (as
is the MCP), this makes the computation more difficult. However, in the
context of SEM this can be seen as less problematic, as equation 3 is
also non-convex.

Additionally, the MCP takes the form of:

\[
pen_{\lambda,\gamma}(\theta_{j}) = \lambda\bigg(|\theta_{j}|-\frac{\theta_{j}^{2}}{2\lambda\gamma}\bigg)I(|\theta_{j}|<\lambda\gamma) +\frac{\lambda^{2}\gamma}{2}I(|\theta_{j}|\geq \lambda\gamma)
\]

with a soft-thresholding update of \[
S(\theta_{j})= 
\begin{cases}
\frac{\gamma}{\gamma-1}S(\theta_{j},\lambda),&  |\theta_{j}| \leq \lambda\gamma\\
\theta_{j} & |\theta_{j}| > \lambda \gamma
\end{cases}
\]

for \(\gamma > 0\). As seen in Figure 2, this results in similar amount
of shrinkage for smaller estimates in comparison to the SCAD, however,
less for larger estimates. For both the SCAD and MCP, both the
\(\gamma\) and \(\lambda\) parameters are used as hyper-parameters. This
involves testing models over a two-dimensional array of parameters,
however, in \textit{regsem}, \(\gamma\) is by default fixed to 3.7 per
\citet{fan2001variable}.

\section{Implementation}\label{implementation}

RegSEM is implemented as the \textit{regsem} package
\citep{jacobucci2016package} in the \textit{R} statistical environment
\citep{statspackage}. To estimate the maximum likelihood fit of the
model, \textit{regsem} uses \textit{Reticular Action Model}
\citep[RAM;][]{McArdle_1984, mcardle2005} notation to derive an implied
covariance matrix. The parameters of each SEM are translated into three
matrices: the \textit{filter} (\textit{F}), the \textit{asymmetric}
(\textit{A}; directed paths; e.g.~factor loadings or regressions), and
the \textit{symmetric} (\textit{S}; undirected paths; e.g.~covariances
or variances). See \citet{jacobucci2016regularized} for more detail on
RAM notation and its application to RegSEM.

Syntax for using the \textit{regsem} package is based on the
\textit{lavaan} package \citep{rosseel2012} for structural equation
models. \textit{lavaan} is a general SEM software program that can fit a
wide array of models with various estimation methods. To use
\textit{regsem}, the user has to first fit the model in \textit{lavaan}.
Note that particularly in cases that the number of variables is larger
than the sample size, the model in lavaan does not need to converge, let
alone run. In this case, the \textbf{do.fit=FALSE} argument in lavaan
can be used. Additionally, regsem only works with models that assume the
variables are continuous, thus none of the additional options in lavaan
that accomodate categorical variables (e.g.~the WLSMV estimator with
ordered indicators) are available.

As a canonical example, below is the code for a confirmatory factor
analysis model with one latent factor and seven indicators from the
\textit{bfi} dataset from the \textit{psych} package
\citep{revelle2014psych}.

\begin{Shaded}
\begin{Highlighting}[]
\KeywordTok{library}\NormalTok{(psych);}\KeywordTok{library}\NormalTok{(lavaan)}
\NormalTok{bfi2 <-}\StringTok{ }\NormalTok{bfi[}\DecValTok{1}\NormalTok{:}\DecValTok{250}\NormalTok{,}\KeywordTok{c}\NormalTok{(}\DecValTok{1}\NormalTok{:}\DecValTok{5}\NormalTok{,}\DecValTok{18}\NormalTok{,}\DecValTok{22}\NormalTok{)]}
\NormalTok{bfi2[,}\DecValTok{1}\NormalTok{] <-}\StringTok{ }\KeywordTok{reverse.code}\NormalTok{(-}\DecValTok{1}\NormalTok{,bfi2[,}\DecValTok{1}\NormalTok{])}

\NormalTok{mod <-}\StringTok{ "}
\StringTok{f1 =~ NA*A1+A2+A3+A4+A5+O2+N3}
\StringTok{f1~~1*f1}
\StringTok{"}
\NormalTok{out <-}\StringTok{ }\KeywordTok{cfa}\NormalTok{(mod,bfi2)}
\CommentTok{#summary(out)}
\end{Highlighting}
\end{Shaded}

After a model is run in lavaan, using \textbf{lavaan()} or any of the
wrapper functions for fitting a model (i.e. \textbf{sem()},
\textbf{cfa()}, or \textbf{growth()}), the object is then used by the
regsem package to translate the model into RAM notation and run using
one of three functions: \textbf{regsem()}, \textbf{multi\_optim()}, or
\textbf{cv\_regsem()}. The \textbf{regsem()} function runs a model with
one penalty value, whereas \textbf{multi\_optim()} does the same but
allows for the use of random starting values. However, the main function
is \textbf{cv\_regsem()}, as this not only runs the model, but runs it
across a vector of varying penalty values. textbf\{cv\_regsem()\} was
originally created to solely use k-fold cross-validation to test
penalties and choose a final model (hence the name). However, as it
currently stands, it is recommended to run the model using the entire
sample, paired with the use of an information criteria to choose a final
model. The use of bootstrapping or k-fold cross-validation requires
additional research and is discussed further in the Discussion.

In the above one-factor model, each of the factor loadings can be tested
with lasso penalties to determine whether each indicator is a necessary
component of the latent factor. The first step is to identify which
parameters are to be penalized, and pass this information to regsem. The
easiest way to accomplish this is through the use of
\textbf{extractMatrices()}:

\begin{Shaded}
\begin{Highlighting}[]
\KeywordTok{library}\NormalTok{(regsem)}
\KeywordTok{extractMatrices}\NormalTok{(out)$A}
\end{Highlighting}
\end{Shaded}

\begin{verbatim}
##    A1 A2 A3 A4 A5 O2 N3 f1
## A1  0  0  0  0  0  0  0  1
## A2  0  0  0  0  0  0  0  2
## A3  0  0  0  0  0  0  0  3
## A4  0  0  0  0  0  0  0  4
## A5  0  0  0  0  0  0  0  5
## O2  0  0  0  0  0  0  0  6
## N3  0  0  0  0  0  0  0  7
## f1  0  0  0  0  0  0  0  0
\end{verbatim}

In this, \textbf{extractMatrices()} allows the user to examine how the
lavaan model is translated into RAM matrices. Further, by looking at the
\textit{A} matrix (directed paths which originate at the column name and
go to the row name), one can identify the parameter numbers
corresponding to the factor loadings of interest for regularization. For
this model, the factor loadings represent parameter numbers one through
seven, of which we pass directly to the \textbf{pars\_pen} argument of
the \textbf{cv\_regsem()} function (if \textbf{pars\_pen=NULL} then all
directed effects, outside of intercepts, are penalized). Additionally,
if parameter labels are used in the lavaan model specification, these
can be directly passed to regsem in the \textbf{pars\_pen} argument.

Additionally, we pass the arguments of how many values of penalty we
want to test (\textbf{n.lambda=15}), how much the penalty should
increase for each model (\textbf{jump=.05}), and finally that lasso
estimation is used (\textbf{type="lasso"}).

\begin{Shaded}
\begin{Highlighting}[]
\NormalTok{out.reg <-}\StringTok{ }\KeywordTok{cv_regsem}\NormalTok{(out, }\DataTypeTok{type=}\StringTok{"lasso"}\NormalTok{, }
                     \DataTypeTok{pars_pen=}\KeywordTok{c}\NormalTok{(}\DecValTok{1}\NormalTok{:}\DecValTok{7}\NormalTok{),}\DataTypeTok{n.lambda=}\DecValTok{23}\NormalTok{,}\DataTypeTok{jump=}\NormalTok{.}\DecValTok{05}\NormalTok{)}
\end{Highlighting}
\end{Shaded}

The \textbf{out.reg} object contains two components,
\textbf{out.reg\$fits} has the parameter estimates for each of the 15
models,

\begin{Shaded}
\begin{Highlighting}[]
\KeywordTok{head}\NormalTok{(}\KeywordTok{round}\NormalTok{(out.reg$parameters,}\DecValTok{2}\NormalTok{),}\DecValTok{5}\NormalTok{)}
\end{Highlighting}
\end{Shaded}

\begin{verbatim}
##      f1 -> A1 f1 -> A2 f1 -> A3 f1 -> A4 f1 -> A5 f1 -> O2 f1 -> N3
## [1,]     0.56     0.77     1.08     0.70     0.90    -0.03    -0.08
## [2,]     0.53     0.74     1.05     0.66     0.87     0.00    -0.05
## [3,]     0.50     0.72     1.03     0.62     0.84     0.00    -0.01
## [4,]     0.47     0.69     1.01     0.58     0.81     0.00     0.00
## [5,]     0.44     0.67     0.99     0.55     0.79     0.00     0.00
##      A1 ~~ A1 A2 ~~ A2 A3 ~~ A3 A4 ~~ A4 A5 ~~ A5 O2 ~~ O2 N3 ~~ N3
## [1,]     1.52     0.69     0.53     1.84     0.88     2.45     2.29
## [2,]     1.53     0.70     0.52     1.84     0.89     2.45     2.29
## [3,]     1.53     0.70     0.52     1.85     0.90     2.45     2.30
## [4,]     1.54     0.71     0.52     1.86     0.90     2.45     2.30
## [5,]     1.55     0.71     0.51     1.87     0.91     2.45     2.30
\end{verbatim}

while \textbf{out.reg\$fits} contains information pertaining to the fit
of each model:

\begin{Shaded}
\begin{Highlighting}[]
\KeywordTok{head}\NormalTok{(}\KeywordTok{round}\NormalTok{(out.reg$fits,}\DecValTok{2}\NormalTok{))}
\end{Highlighting}
\end{Shaded}

\begin{verbatim}
##      lambda conv rmsea     BIC
## [1,]   0.00    0  0.08 5713.57
## [2,]   0.05    0  0.08 5708.76
## [3,]   0.10    0  0.08 5710.45
## [4,]   0.15    0  0.08 5707.20
## [5,]   0.20    0  0.09 5709.89
## [6,]   0.25    0  0.09 5713.11
\end{verbatim}

In this, the user can examine the penalty (lambda), whether the model
converged (``conv''=0 means converged, whereas either 1 or 99 is
non-convergence), and the fit of each model. By default, two fit indices
are output, both the root mean square error of approximation
\citep[RMSEA;][]{steiger1980}, and the Bayesian information criteria
\citep[BIC;][]{schwarz1978estimating}. Both the RMSEA and BIC take into
account the degrees of freedom of the model, an important point for
model selection in the presence of lasso penalties (and other penalties
that set parameters to zero). \citet{Zou2007} proved that the number of
nonzero coefficients is an unbiased estimate of the degrees of freedom
for regression. As the penalty increases, select parameters are set to
zero, thus increasing the degrees of freedom, which for fit indices that
include the degrees of freedom in the calculation, means that although
\(F_{ML}\) may only get worse (increase), both the RMSEA and BIC can
improve (decrease).

Instead of examining the \textbf{out.reg\$fits} output matrix of
parameter estimates, users also have the option to plot the trajectory
of each of the penalized parameters. This is accomplished with

\begin{Shaded}
\begin{Highlighting}[]
\KeywordTok{plot}\NormalTok{(out.reg,}\DataTypeTok{show.minimum=}\StringTok{"BIC"}\NormalTok{)}
\end{Highlighting}
\end{Shaded}

\includegraphics{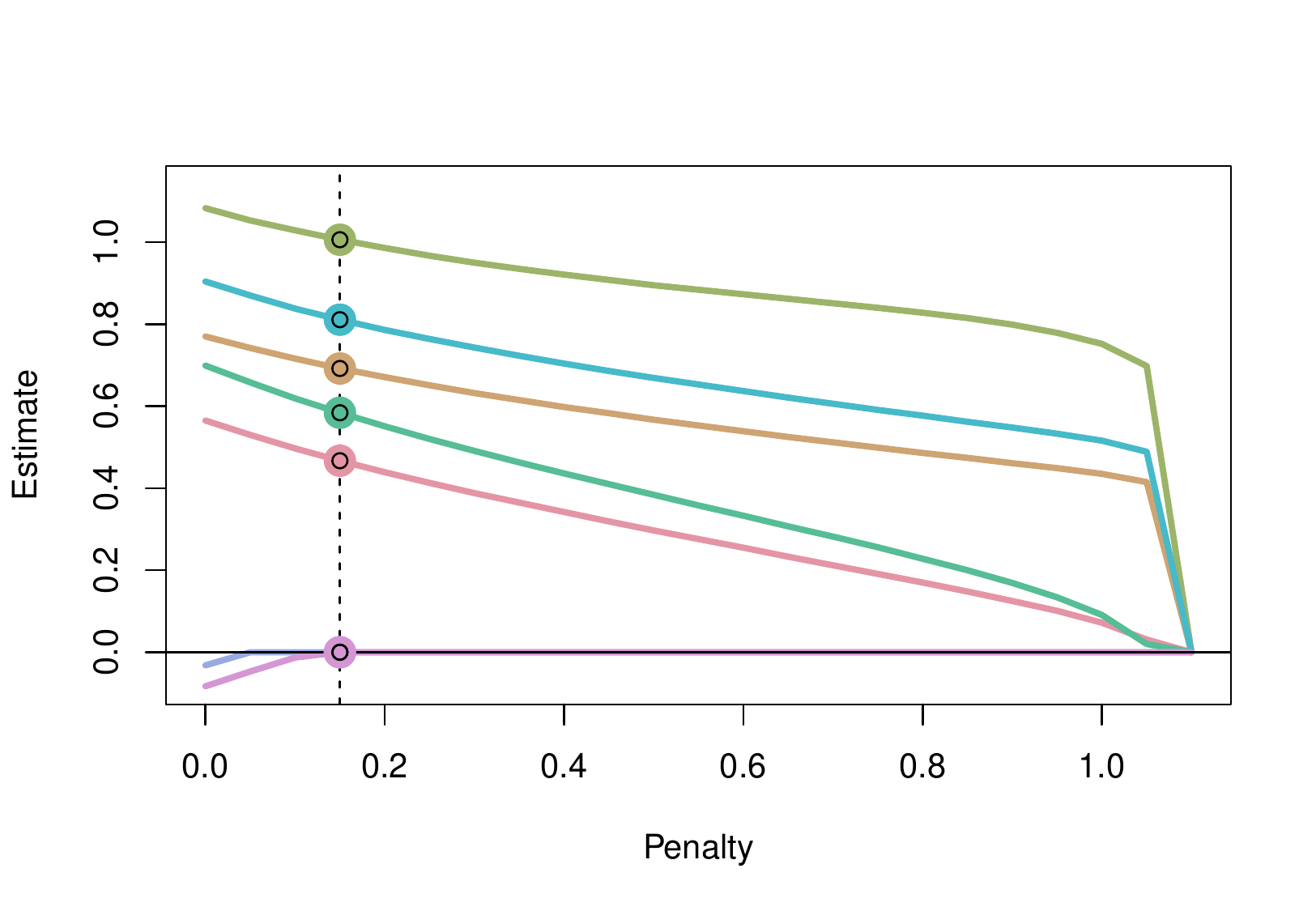}

After a final model (penalty) is chosen, users have the option either
just use the output from \textbf{cv\_regsem()}, or the final model can
be re-run with either \textbf{regsem()} or \textbf{multi\_optim()} to
attain additional information. In the model above, the best fitting
penalty, according to the BIC, is \(\lambda=0.15\).

\begin{Shaded}
\begin{Highlighting}[]
\KeywordTok{summary}\NormalTok{(out.reg)}
\end{Highlighting}
\end{Shaded}

\begin{verbatim}
## CV regsem Object
##  Number of parameters regularized: 7
##  Lambda ranging from 0 to 1.1
##  Lowest Fit Lambda: 0.15
##  Metric: BIC
##  Number Converged: 23
\end{verbatim}

Instead of having to re-run the model with \textbf{regsem()} to get the
final parameter estimates, a user can specify what fit index should be
used to choose a final model with \textbf{metric = }in
\textbf{cv\_regsem()}. These estimates are printed in:

\begin{Shaded}
\begin{Highlighting}[]
\NormalTok{out.reg$final_pars}
\end{Highlighting}
\end{Shaded}

\begin{verbatim}
## f1 -> A1 f1 -> A2 f1 -> A3 f1 -> A4 f1 -> A5 f1 -> O2 f1 -> N3 A1 ~~ A1 
##    0.467    0.692    1.006    0.584    0.811    0.000    0.000    1.540 
## A2 ~~ A2 A3 ~~ A3 A4 ~~ A4 A5 ~~ A5 O2 ~~ O2 N3 ~~ N3 
##    0.707    0.515    1.859    0.903    2.448    2.298
\end{verbatim}

Additional fit indices can be attained through the
\textbf{fit\_indices()} function if only one model was run with either
\textbf{regsem()} or \textbf{multi\_optim()}. These same fit measures
can be accessed through \textbf{cv\_regsem()} through changing the
defaults with the \textbf{fit.ret=c("rmsea","BIC")} argument. Finally,
instead of assessing these fit indices on the same sample that the
models were run on, a holdout dataset could be used. This can be done
two ways: either with \textbf{cv\_regsem(...,fit.ret2="test")} or with
\textbf{fit\_indices(model,CV=TRUE,CovMat=)} and specifying the name of
the holdout covariance matrix.

Structural equation modeling is hard, and pairing with regularization
doesn't make it any easier. Given this, and the number of options
available in the \textit{regsem} package, a Google group forum was
created in order to answer questions and trouble shoot at
\url{https://groups.google.com/forum/#!forum/regsem}.

\section{Comparison}\label{comparison}

To compare the different types of penalties in \textit{regsem}, we
return to the the initial example of the latent growth curve model
displayed in Figure 1. Using the same simulated data, the model can be
run in \textit{lavaan} as

\begin{Shaded}
\begin{Highlighting}[]
\NormalTok{mod1 <-}\StringTok{ "}
\StringTok{i =~ 1*x1 + 1*x2 + 1*x3 + 1*x4}
\StringTok{s =~ 0*x1 + 1*x2 + 2*x3 + 3*x4}
\StringTok{i ~ c1 + c2 + c3 + c4 + c5 + c6 + c7 + c8 + c9 + c10}
\StringTok{s ~ c1 + c2 + c3 + c4 + c5 + c6 + c7 + c8 + c9 + c10}
\StringTok{"}
\NormalTok{lav.growth <-}\StringTok{ }\KeywordTok{growth}\NormalTok{(mod1,dat,}\DataTypeTok{fixed.x=}\NormalTok{T)}
\end{Highlighting}
\end{Shaded}

Comparing different types of penalties in \textit{regsem} requires a
different specification of the \textbf{type} argument. The options
currently include maximum likelihood (\textbf{"none"}), ridge
(\textbf{"ridge"}), lasso (\textbf{"lasso"}; the default), adaptive
lasso (\textbf{"alasso"}), elastic net (\textbf{"enet"}), SCAD
(\textbf{"scad"}), and MCP (\textbf{"mcp"}). For the elastic net, there
is an additional hyperparameter, \textbf{alpha} that controls the
tradeoff between ridge and lasso penalties. This is specified as
\textbf{alpha=} , which has a default of 0.5. Additionally, both the
SCAD and MCP have the additional hyper parameter of \textbf{gamma},
which is specified as \textbf{gamma=} and defaults to 3.7 per
\citet{fan2001variable}.

For the purposes of comparison, each of the 20 covariate regressions
were penalized using the lasso, adaptive lasso, SCAD, and MCP, and
compared to the maximum likelihood estimates. In this model, the data
were simulated to have two large effects (both \textbf{c1} parameters),
two small effects (both \textbf{c2} parameters) and sixteen true zero
effects (\textbf{c3-c10} parameters). Note that the covariates were
simulated to have zero covariance among each variable. If there was
substantial collinearity among covariates, the elastic net would be more
appropriate to simultaneously select predictors while also accounting
for the collinearity. The parameter estimates corresponding the the best
fit of the BIC are has the fit of each model, resulting in Table 1,
created using the \textit{xtable} package \citep{dahl2009xtable}.

\begin{table}[ht]
\centering
\begin{tabular}{rrrrrr}
  \hline
 & MLE & lasso & alasso & SCAD & MCP \\ 
  \hline
$c1_{i}$ & 0.92* & 0.72 & 0.91 & 0.94 & 0.92 \\ 
  $c2_{i}$ & 0.07 & 0.00 & 0.00 & 0.00 & 0.00 \\ 
  $c3_{i}$ & 0.10 & 0.00 & 0.00 & 0.00 & 0.00 \\ 
  $c4_{i}$ & 0.07 & 0.00 & 0.00 & 0.00 & 0.00 \\ 
  $c5_{i}$ & 0.04 & 0.00 & 0.00 & 0.00 & 0.00 \\ 
  $c6_{i}$ & -0.25 & 0.00 & 0.00 & 0.00 & -0.19 \\ 
  $c7_{i}$ & 0.11 & 0.00 & 0.00 & 0.00 & 0.00 \\ 
  $c8_{i}$ & -0.13 & 0.00 & 0.00 & 0.00 & 0.00 \\ 
  $c9_{i}$ & -0.03 & 0.00 & 0.00 & 0.00 & 0.00 \\ 
  $c10_{i}$ & 0.09 & 0.00 & 0.00 & 0.00 & 0.00 \\ 
  $c1_{s}$ & 1.18* & 1.09 & 1.22 & 1.24 & 1.24 \\ 
  $c2_{s}$ & 0.29* & 0.19 & 0.28 & 0.35 & 0.35 \\ 
  $c3_{s}$ & 0.18 & 0.09 & 0.00 & 0.00 & 0.00 \\ 
  $c4_{s}$ & -0.08 & 0.00 & 0.00 & 0.00 & 0.00 \\ 
  $c5_{s}$ & -0.18 & 0.00 & 0.00 & 0.00 & 0.00 \\ 
  $c6_{s}$ & 0.25* & 0.00 & 0.00 & 0.00 & 0.00 \\ 
  $c7_{s}$ & -0.18 & -0.04 & 0.00 & 0.00 & 0.00 \\ 
  $c8_{s}$ & 0.26* & 0.10 & 0.00 & 0.00 & 0.00 \\ 
  $c9_{s}$ & -0.06 & 0.00 & 0.00 & 0.00 & 0.00 \\ 
  $c10_{s}$ & 0.08 & 0.00 & 0.00 & 0.00 & 0.00 \\ 
  BIC & 3465.28 & 3427.46 & 3415.05 & 3414.38 & 3417.20 \\ 
   \hline
\end{tabular}
\caption{Parameter estimates for the final models across five estimation methods. Note that * represent significant parameters at p < .05 for maximum likelihood estimation.}
\end{table}

While every regularization method erroneously set both simulated true
intercept effects as zero (non-significant in MLE), both the adaptive
lasso and SCAD correctly identified every true zero effect. The lasso
identified two false effects while the MCP mistakenly identified one.
Additionally, the lasso estimation of the true effects was attentuated
in comparison to the other regularization methods. This is in line with
previous research \citep{fan2001variable}, necessitating the use of a
two-step relaxed lasso method
\citetext{\citealp{meinshausen2007relaxed}; \citealp[see][]{jacobucci2016regularized}}
As expected given the small ratio between number of estimated parameters
and sample size, MLE mistakenly identified 3 false effects as
significant.

To compare the performance of each penalization method further,
particularly in the presence of a small parameter to sample size ratio,
a small simulation study was conducted. The same model and effects was
kept, but the sample size was varied to include 80, 200, and 1000 to
demonstrate how MLE improves as sample size increases, while each of the
regularization methods performs well regardless of sample size. Each run
was replicated 200 times. For each regularization method, the BIC was
used to choose a final model among the 40 penalty vales. The results are
displayed in Table 2.

\begin{table}[ht]
\centering
\begin{tabular}{rrrrrrr}
  \hline
 & N & ML & lasso & alasso & SCAD & MCP \\ 
  \hline
 & 80.00 & 0.08 & 0.08 & \textbf{0.04} & 0.05 & 0.20 \\ 
  False Positives & 200.00 & 0.06 & 0.05 & \textbf{0.02} & 0.03 & 0.06 \\ 
   & 1000.00 & 0.05 & 0.08 & \textbf{0.01} & \textbf{0.01} & 0.02 \\ 
   & 80.00 & 0.33 & \textbf{0.31} & 0.35 & 0.35 & 0.31 \\ 
  False Negatives & 200.00 & \textbf{0.19} & \textbf{0.19} & 0.23 & 0.26 & 0.27 \\ 
   & 1000.00 & \textbf{0.00} & \textbf{0.00} & \textbf{0.00} & 0.01 & 0.03 \\ 
   \hline
\end{tabular}
    \caption{Results from the simulation using the model in Figure 1. Each condition was replicated 200 times. False positives represent concluding that the simulated regressions of zero were concluded as nonzero. False negatives are concluding that either the simulated regression values of 1 or 0.2 are in fact zero. Bolded values represent the smallest error per condition.}
\end{table}

For false positives, the adaptive lasso demonstrated the best
performance, where the performance of MLE leveled off at the 0.05 level
at a sample size of 1000 as expected. For false negatives, lasso
penalties demonstrated similar results to MLE. This was expected given
the tendency of the lasso to under-penalize small coefficients in
comparison to the other regularization methods. The adaptive lasso and
SCAD demonstrated slightly worse results, however, outside of the MCP,
each method made either zero or near zero errors at a sample size of
1000. The poor performance of the MCP may be in part due to fixing the
\(\gamma\) penalty to 3.7. Varying this parameter may improve the
performance of the method. In summary, the regularization methods
demonstrated an improvement over maximum likelihood, particularly at
small samples, for a model that had a large number of estimated
parameters.

\section{Discussion}\label{discussion}

This paper provides an introduction to the \textit{regsem} package,
outlining the mathematical details of regularized structural equation
modeling \citep[RegSEM;][]{jacobucci2016regularized} and the usage of
the \textit{regsem} package. RegSEM allows the use of regularization
while keeping the structural equation model intact, adding penalization
directly into the estimation of the model. The application of RegSEM was
detailed using two example models: a latent growth curve model with 20
predictors of both the latent intercept and slope, along with a factor
analysis with one latent factor. With the latent growth curve model, the
small parameter to sample size ratio resulted in a larger number of
false positives in using maximum likelihood estimation. In both the
simulated example and the small simulation, the different types of
regularization in \textit{regsem} demonstrated better false positive and
negative rates in comparison to maxiumum likelihood across sample sizes.

Broadly speaking, there is a growing amount of research into the
integration between data mining methods and latent variable models.
Specifically, beyond RegSEM, this has taken the form of item response
theory and regularization \citep{sun2016latent}, other regularization
and latent variable formulations \citep{fanc, huang2017penalized},
pairing both structural equation models with decision trees
\citep{brandmaier2013}, exploratory psychological network analaysis
\citep[e.g.][]{epskamp2016generalized}, along with many others. The
amount of pairing between methods that have generally been housed in
separate camps will only increase into the future. This type of research
will be facilitated by the general upsurge in the creation of open
source software that gives users a general framework to test models.
This was the motivation behind creating the \textit{regsem} package, in
that users can estimate models ranging from simple factor analysis
models, to latent longitudinal models with few to many time points, and
finally to models with a large number of latent and observed variables.
The use of regularization allows for the estimation of much larger
structural equation models than before. However, sample sizes in the
social and behavioral sciences are typically not large. To estimate
large models with small sample sizes invites increasing amounts of bias
as demonstrated with the simulated data in this paper. Regularization
can be used to reduce the complexity of the model, thus decreasing both
the bias and variance.

With highly constrained structural equation models, achieving model
convergence can be particularly problematic in using \textit{regsem}.
For instance, with the latent change score model
\citep{mcardle2001latent}, Bayesian regularization methods have less
difficulty in reaching convergence across chains
\citep{jacobucci2017JackChapter}. With the recent advent of additional
sparsity inducing priors, along with new forms of software such as Stan
\citep{carpenter2016stan}, for some models it may be more appropriate to
use these Bayesian regularization methods over their frequentist
counterparts. In the realm of Bayesian regularization for structural
equation models, although some research exists \citep{feng2017bayesian},
much more is warranted.

Future research with \textit{regsem} should focus on a number of
avenues. One is comparing the different forms of regularization,
delineating which method may be best in which setting. Additionally, as
structural equation models become larger, with the advent of much larger
datasets, computational speed will become a principal concern. Although
40 penalties in the models tested above can be run in a matter of
seconds on a standard laptop, larger models can take much longer. To
handle this, future implementation with \textit{regsem} will test the
inclusion of different types of optimization, specifically testing
whether coordinate descent algorithms \citep{friedman2010regularization}
can speed up convergence. Finally, although the use of bootstrapping or
k-fold cross-validation is computationally intensive, these forms of
resampling (paired with a fit index that does not have a penalty for the
number of parameters, i.e. \(\chi^{2}\)) may produce better results for
both ridge and elastic net penalties, where it is less clear how to take
into account parameter shrinkage for choosing a final model.

\subsection{Conclusion}\label{conclusion}

This paper provided a brief overview on the use of the \textit{regsem}
package as an implementation of regularized structural equation
modeling. Because structural equation modeling encompasses a wide array
of latent variable models, the \textit{regsem} package was created as a
general package for including different forms of regularization into a
host of latent variable models. RegSEM, and thus the \textit{regsem}
package, has been evaluated in a wide array of SEM models, including
confirmatory factor analysis \citep{jacobucci2016regularized}, latent
change score models \citep{jacobucci2017JackChapter}, and mediation
models \citep{serang2017xmed}. Future updates to \textit{regsem} will
focus on decreasing the computational time of large latent variable
models in order to provide an avenue of testing for researchers
collecting larger and larger datasets. RegSEM is a method that operates
at all ends of the data size spectrum: allowing for a reduction in
complexity when the sample size is small, along with dimension reduction
in the presence of large data (both \(N\) and \(P\)).

\renewcommand\refname{References}
\bibliography{jacobucci}

\end{document}